\documentstyle[aps,tighten]{revtex}
\begin{document}
\input{epsf}
 \draft

\title{Vacuum Polarization and Energy Conditions at a Planar Frequency
       Dependent Dielectric to Vacuum Interface}

\author{\bf Michael J. Pfenning\footnote{email: mitchel@physics.uoguelph.ca}}
\address{Department of Physics, University of Guelph,
         Guelph, Ontario, N1G 2W1, Canada}

\date{January 19, 2000}

\maketitle

\begin{abstract}
The form of the vacuum stress-tensor for the quantized scalar field at
a dielectric-vacuum interface is studied.  The dielectric is modeled
to have an index of refraction that varies with frequency. We find 
that the stress-tensor components, derived from the mode function
expansion of the Wightman function, are naturally regularized by the
reflection and transmission coefficients of the mode at the boundary.
Additionally, the divergence of the vacuum energy associated with a
perfectly reflecting mirror is found to disappear for the  dielectric
mirror at the expense of introducing a new energy density near
the surface which has the opposite sign.  Thus the weak energy condition
is always violated in some region of the spacetime. For the dielectric
mirror, the mean vacuum energy density per unit plate area in a
constant time hypersurface is always found to be positive (or zero) and
the averaged weak energy condition is proven to hold for all observers
with non-zero velocity along the normal direction to the boundary.
Both results are found to be generic features of the vacuum stress-tensor
and not necessarily dependent of the frequency dependence of the dielectric.
\end{abstract}


\section{Introduction}
The renormalized vacuum stress-tensor for quantized fields
at perfectly reflecting boundaries has been extensively studied,
sufficiently so that they have made their way into various texts.
Fulling \cite{Fulling} and Birrell and Davies \cite{Brl&Dv} are
two examples.  The perfect reflector enforces a Dirichlet boundary
condition that must be satisfied by the quantized field.  For the
case of the scalar or electromagnetic field with an infinite planar
boundary, the renormalized vacuum energy density for stationary
observers has the form
\begin{equation}
\langle \rho \rangle_{\mbox{\tiny Ren.}} = {\alpha \over x^m},
\end{equation}
where $\alpha$ is a unitless constant of order unity, $x$ is the
transverse distance from the planar boundary and $m > 1$ is the
dimension of the spacetime.  The most notable property of the
energy density is the divergence in the limit $x\rightarrow 0$.
 
The divergence is not at all unexpected.  The application of
the Dirichlet boundary condition fixes the value of the field at a
specific point with no uncertainty in the position.  Thus the
uncertainty of the conjugate momenta must be infinite, giving
rise to the stress-tensor divergence.  The precise mathematical
origin of the divergence comes from the renormalization of the
vacuum energy via the Green's function.  The argument
is simply that the unrenormalized Green's function 
is constructed such that it vanishes on the boundary.  To renormalize
the Green's function, the corresponding Minkowski space
Green's function must be subtracted away.  Since this is divergent
in the limit $x\rightarrow 0$, the subtraction carried out for 
renormalization will leave behind a divergent part.   When the vacuum
energy is calculated from the renormalized Green's function, it 
will diverge at $x=0$ also.

The divergence of the vacuum energy density causes considerable
consternation since the vacuum would contain an infinite amount of
energy. In a self consistent theory of gravity, the back reaction
of the vacuum energy on the spacetime would be considerable.  Yet we
know from experience that metallic foils do not cause significant
gravitational distortions around themselves.

There are three resolutions to the problems: (1) Replace the Dirichlet 
boundary condition with something more realistic, which is the topic of
this paper.  (2) Add position uncertainty to the mirror;
this has been discussed by Ford and Svaiter \cite{F&SV97b}.  (3) Treat
the entire problem quantum mechanically, which has been partially touched
upon by Barton and Eberlein \cite{B&EB93} and more recently by Helfer
\cite{Helfer}.  

Replacing the Dirichlet boundary condition has to some degree been
discussed by other authors.  For example Candelas \cite{Candelas}
has derived the asymptotic form of the vacuum energy density for the
problem of two nondispersive dielectric half-spaces which are in contact
with one another along a planar interface.  More recently Helfer and
Lang \cite{He&La98} carried out similar calculations for a
nondispersive dielectric to vacuum interface, with considerable interest
placed upon the energy conditions in the vacuum region.  

This paper addresses the behavior of the stress-tensor for a
quantized massless scalar field at a planar dielectric to vacuum
interface where the index of refraction of the material varies as
a function of frequency.  The Dirichlet boundary condition is 
replaced with appropriate frequency dependent reflection
and transmission coefficients.  The
stress-tensor in the vacuum region is well behaved, even on the
interface, and is largely independent of the functional dependence
of the index of refraction.   Therefore the vacuum energy divergence
is a result of the Dirichlet boundary condition and not a generic
feature of the field theory.  It should be noted, however, that it is
possible to get divergences for dielectric materials at the interface.
The simplest example is a material with a constant index of refraction.
In this case, the divergence comes about because part of the amplitude
of the wave at every frequency still satisfies Dirichlet boundary
conditions.  The existence of the divergence is intimately tied to 
the failure of specific integrals to exist. For example, in two
dimensions, a divergence will occur if 
\begin{equation}
\int_0^\infty R(\omega)\, d\omega
\label{eq:div_cond}
\end{equation}
fails to exist.  Here $R(\omega) = \left[n(\omega)-1\right]/\left[n(\omega)
+1\right]$ is the reflection coefficient for the modes at the interface,
while $n(\omega)$ is the index of refraction. There
are similar integrals in four dimensions involving both the reflection and
transmission coefficients.   

There are two remarkable results regarding frequency dependent
dielectric models.
First, all of the integrals that must be evaluated to find the components
of the stress-tensor are self regularized, meaning that they all have a
built in cutoff function that prevents ultraviolet divergences. 
For example, in two dimensions the energy density for a static observer is
\begin{equation}
\langle \rho \rangle _{\mbox{\tiny Ren.}} = -{2\xi\over\pi} 
\int_{0}^{\infty} \omega \,R(\omega)\, \cos(2\omega x) \, d\omega .
\label{eq:2D_result}
\end{equation}
The only constraint placed on the index of
refraction is for it to tend to unity as the frequency grows.  Thus
the reflection coefficient plays the role of the regulator in the
integral.  Similar results are found for the four-dimensional case
as well, where there exists another integral for the evanescent modes
which is regularized by the transmission coefficient.  

The other unique property, which we shall prove in general, is that the 
vacuum stress-tensor satisfies the averaged weak energy condition (AWEC) 
for any moving geodesic observer, provided that the component of the
observer's four velocity  normal to the dielectric to vacuum
interface is non-zero. This occurs even though the
pointwise weak energy condition (WEC) over some regions of the
spacetime fails.  The AWEC results from the Riemann-Lebesgue lemma
applied to integrals of the form of Eq.~(\ref{eq:2D_result}) 
evaluated in the  limit $x\rightarrow\infty$.  In order to apply
the lemma, it is required that conditions like (\ref{eq:div_cond})
exist.

A related result is that the energy density, integrated along
the normal direction in a constant time hypersurface
in the vacuum region, is a positive constant. In two dimensions
the constant is zero and independent of the choice of model for
the index of refraction.  Thus the spacetime always contains
two regions, one where the vacuum energy is positive and another
with equal negative energy.  In four dimensions, the constant is
found by evaluating the integral
\begin{equation}
{E \over \mbox{unit area}} = {1\over 64\pi} \int_0^\infty \omega^2
\left[n^2(\omega) - 1\right] d\omega
\end{equation}
for a specific choice of $n(\omega)$.  We consider only the class
of models for which $n(\omega) \geq 1$ for all frequencies.
The function space for which the integral holds is obviously much
larger, however this is beyond the focus of this paper.    

Specific models are examined in this paper.  The vacuum stress-tensor
is found to asymptotically approach the form of the perfectly
reflecting vacuum stress-tensor in the large distance limit.  The 
price of removing the divergence of the stress-tensor by using a
frequency dependent  dielectric is the introduction of a vacuum
energy near the surface with opposite sign.  This simultaneously
is the reason why the AWEC holds while the WEC fails.

\section{Mode Structure at a Dielectric Interface}\label{sec:modes}

We begin with the source free scalar wave equation in m+1 dimensional
Minkowski spacetime for a nondispersive material,
\begin{equation}
\left(\nabla^2 - n^2 \partial_t^2\right)\phi({\bf x},t) = 0,
\label{eq:WE_ordinary}
\end{equation}
where $n$ is a constant. Units of $\hbar = c = 1$ are being used throughout.
A positive frequency plane wave solution to this
equation in an infinite medium takes the form
\begin{equation}
f_{\bf k}({\bf x},t) \propto e^{i(n{\bf k}\cdot{\bf x}- \omega t)},
\end{equation}
where $\omega = |{\bf k}|$. The modes can be
normalized by defining the probability density for the wave equation
and setting its integral over all space to unity.  This
leads to the normalization condition
\begin{equation}
\left( f_{\bf k}({\bf x},t) , f_{\bf k'}({\bf x},t)\right) \equiv
-i\int d^mx \, n^2 \left\{ f_{\bf k}({\bf x},t) \left[\partial_t
f_{\bf k'}^*({\bf x},t)\right] - \left[\partial_t f_{\bf k}({\bf x},t)
\right] f_{\bf k'}^*({\bf x},t) \right\} = \delta^m ({\bf k}-{\bf k'}).   
\end{equation}
Thus the normalized plane wave mode takes the form
\begin{equation}
f_{\bf k}({\bf x},t) = n^{(m-2)/2}\,\left[ 2\omega (2\pi)^m \right]
^{-1/2}e^{i(n{\bf k}\cdot{\bf x}- \omega t)}
\end{equation}
The most general solution to the wave equation~(\ref{eq:WE_ordinary})
is then given as a mode function expansion,
\begin{equation}
\phi({\bf x},t) = \int_{-\infty}^{+\infty} d^m k \left[ a_{\bf k}
 f_{\bf k}({\bf x},t) +  a_{\bf k}^\dagger  f_{\bf k}^*({\bf x},t)\right].
\label{eq:mode_expansion}
\end{equation}
If $\phi({\bf x},t)$ is a classical solution to the wave equation,
then $a_{\bf k}$ and $a_{\bf k}^\dagger$ are the Fourier coefficients
for the mode function expansion.  The classical solution
has the property that it can in general be written as
\begin{equation}
\phi({\bf x},t) = g(x-{1\over n}t) + h(x+ {1\over n}t).
\end{equation}
This represents two traveling waves, $g$ and $h$, that propagate
in the $+\hat{x}$ and $-\hat{x}$ directions, respectively.  Each traveling
wave maintains a constant spatial profile that is just translated
along the direction of motion.  When second quantization is applied
the coefficients $a_{\bf k}$ and $a_{\bf k}^\dagger$ become the
annihilation and creation operators, respectively. 

Next, consider wave propagation in a dispersive material.  A 
scalar wave equation is needed which mimics the behavior of the
electromagnetic field inside a dispersive material.  In 
electromagnetism, the wave equation is found by first Fourier
expanding the Maxwell equations for harmonic time dependence.  Upon
combining them, we find that each component of $\bf E$ and $\bf B$ must
individually satisfy the Helmholtz wave equation 
\begin{equation}
\left[\nabla^2 - n^2(\omega) \omega^2\right]\phi({\bf x},\omega) = 0.
\label{eq:WE_Helmholtz}
\end{equation}
We will take this to be our dispersive wave equation.  The mode
functions which satisfy this equation are the same as given above
with the replacement $n \rightarrow n(\omega)$.  The general solution
$\phi({\bf x},t)$ is given as a mode function expansion in terms
of the plane waves in the same form as (\ref{eq:mode_expansion}).
To obtain this form  we must impose the condition $n^*(\omega) = 
n(\omega)$.  However, even a simple resonance model (anomalous
dispersion) for the dielectric as given in Jackson \cite{Jackson}
leads to an index of refraction that does not satisfy this condition
and would necessitate a more complicated quantum field theory.  Thus,
we restrict $n(\omega)$ to be real for simplicity.  The physical
interpretation of this condition is that the dielectric does not
attenuate the wave propagating through it. It also excludes materials
that may have any degree of conductivity.  Strictly speaking, causality
and the Kramers-Kronig relations require any index of refraction to
have a nonzero imaginary part. However, it is also possible to have
broad frequency bands within which the real part is very large compared
to the imaginary part.  Thus, for the models studied here, we are
assuming that the dominant contribution to the
energy density come from regions of the spectrum where $Re(n)\gg Im(n)$.
The general solution to
the wave equation in the time domain no longer has the physical
property of traveling waves as in the nondispersive material.
For this case, each frequency component propagates with a different
velocity, so the overall profile of the wave changes as it moves.

Now consider the example of two different dielectrics which have a
planar interface at $x_1=0$. To the left of the planar boundary the
index of refraction will be denoted by $n_I$.  To the right of the
boundary the index will be $n_T$.  In addition we will define the
normal to the interface, $\hat{\bf u}= (1,0,\cdots,0)$.
 We allow a monochromatic plane wave to be incident on the
interface from the left with a momentum ${\bf k}_I$.  At the interface
this incident wave is partly reflected back into the first medium and
partly transmitted into the second medium.  The
reflected and transmitted waves have momenta ${\bf k}_r$ and
${\bf k}_t$ respectively, and the three components of the wave are
\begin{eqnarray}
\phi_{inc. }({\bf x},t) &=&     e^{i(n_I{\bf k}_I\cdot{\bf x}- \omega_I t)}\\
\phi_{refl.}({\bf x},t) &=& R\, e^{i(n_I{\bf k}_R\cdot{\bf x}- \omega_R t)}\\
\phi_{tran.}({\bf x},t) &=& T\, e^{i({\bf k}_T\cdot{\bf x}- \omega_T t)}
\end{eqnarray}
For electromagnetism, the plane wave modes at the boundary would
have to satisfy the continuity of the transverse and perpendicular
components of the fields across the interface.  For the scalar
field, the boundary conditions that will be applied are the
continuity of $\phi({\bf x},t)$ and $\partial_x \phi({\bf x},t)$
across the boundary.  This leads to several relations:
\begin{enumerate}

\item	The continuity relations must hold at all times, therefore
	we must impose 
	\begin{equation}
	\omega_I = \omega_R = \omega_T =\omega.
	\end{equation} 

\item	All of the terms must have the same functional dependence
        on the surface of the interface, therefore
        \begin{equation}
        n_I{\bf k}_I\cdot{\bf x}\,=\,n_R{\bf k}_R\cdot{\bf x}=
        {\bf k}_T\cdot{\bf x} \;\;\mbox{at}\; x_1=0.
        \end{equation}
        Taking the first two of these relations yields the 
        law of reflection, $\theta_I = \theta_R$.  Taking
        the first and third of these relations yields Snell's
        law, $n_I \sin \theta_I = n_T \sin \theta_T$. This
        allows us to solve for the components of ${\bf k}_R$
        and ${\bf k}_T$ in terms of the components of ${\bf k}_I$,
        \begin{eqnarray}
        {\bf k}_I &=& ( k_1, k_2, \cdots , k_m),\\
        {\bf k}_R &=& (-k_1, k_2, \cdots , k_m),\\
        {\bf k}_T &=& (\beta, n_I k_2, \cdots , n_I k_m), 
        \end{eqnarray} 
	where
	\begin{equation}
	\beta = \sqrt{ n_T^2 \omega^2 - n_I^2\left(k_2^2+\cdots+k_m^2
	\right)}.
	\end{equation}
	For $n_I>n_T$, it is evident that $\beta$ will be imaginary for
	some set of incident momenta.  These are the evanescent modes
	for which the incident wave undergoes total internal reflection.
	At the interface, all of the incident energy will be reflected
	while there exists an exponentially decaying tail in the medium
	that has a lower index of refraction. This theory of  scalar
	dielectrics simulates all of the properties for wave
	reflection and transmission of the electromagnetic field.
	
\item	The result from the continuity relations are the
	reflection coefficient,
	\begin{equation}
	R = {{k_1 - \sqrt{ \left({n_T\over n_I}\right)^2\omega^2 -
	\left(k_2^2+\cdots+k_m^2\right)}} \over
	{k_1 + \sqrt{ \left({n_T\over n_I}\right)^2\omega^2 -
	\left(k_2^2+\cdots+k_m^2\right)}}},
	\end{equation}
	and the transmission coefficient
	\begin{equation}
	T = {{2k_1} \over
	{k_1 + \sqrt{ \left({n_T\over n_I}\right)^2\omega^2 -
	\left(k_2^2+\cdots+k_m^2\right)}}}.
	\end{equation}
	These relations are identical to the reflection and transmission
	coefficients for the transverse magnetic (TM) modes of the
	electromagnetic field.
\end{enumerate}
The mode functions can now be easily written down.  There are 
two types, ``right-going'' which have $k_1>0$ and ``left-going''
which have $k_1<0$.  The right-going modes have the form
\begin{equation}
f_{\bf k}({\bf x},t) ={ n_I^{(m-2)/2}\over\left[ 2\omega (2\pi)^m
\right]^{1/2}} \left\{ \Theta(-x) \left[ e^{in_Ik_1x_1} + R_{rg}
e^{-in_Ik_1x_1}\right]+ \Theta(x) T_{rg} e^{i\beta x_1} \right\}
e^{in_I(k_2x_2+\cdots +k_mx_m)} e^{-i\omega t},
\label{eq:Gen_right_mode}
\end{equation}
where $\Theta(x)$ is the step function
\begin{equation}
\Theta(x) = \left\{ \begin{array}{cl}
                 0 & \mbox{for } x\leq 0,\\ 
                 1 & \mbox{for }x>0.
              \end{array}\right.
\end{equation}
In addition the right-going reflection and transmission coefficients
are $R_{rg} = R$ and $T_{rg} = T$ as defined above.

The left going mode functions have a similar form,
\begin{equation}
f_{\bf k}({\bf x},t) ={ n_T^{(m-2)/2}\over\left[ 2\omega (2\pi)^m
\right]^{1/2}} \left\{ \Theta(-x) T_{lg} e^{-i\beta' x_1} +
\Theta(x) \left[ e^{in_Tk_1x_1} + R_{lg} e^{-in_Tk_1x_1}\right]
\right\} e^{in_T(k_2x_2+\cdots +k_mx_m)}e^{-i\omega t},
\label{eq:Gen_left_mode}
\end{equation}
with left-going reflection and transmission coefficients given
by the interchange of $n_I$ with $n_T$.  In addition, the change
in sign of the x-component of the momentum must be accounted for.
Thus the left-going reflection coefficient is
\begin{equation}
R_{lg} = {{-k_1 - \sqrt{ \left({n_I\over n_T}\right)^2\omega^2 -
\left(k_2^2+\cdots+k_m^2\right)}} \over
{-k_1 + \sqrt{ \left({n_I\over n_T}\right)^2\omega^2 -
\left(k_2^2+\cdots+k_m^2\right)}}},
\end{equation}
and the transmission coefficient
\begin{equation}
T_{lg} = {{-2k_1} \over
{-k_1 + \sqrt{ \left({n_I\over n_T}\right)^2\omega^2 -
\left(k_2^2+\cdots+k_m^2\right)}}}.
\end{equation}
In addition, the new propagation constant is
\begin{equation}
\beta' = \sqrt{ n_I^2 \omega^2 - n_T^2\left(k_2^2+\cdots+k_m^2\right)}.
\end{equation}

The normalization of the mode functions has been chosen such
that the incident plane wave portion of each mode satisfies the
orthogonality relation in an infinite medium. We are now prepared
to begin our calculation of the vacuum stress-tensor outside
of a frequency dependent dielectric material.


\section{Two-dimensional dielectric half-space}

\subsection{Vacuum stress-tensor}

We begin with a two-dimensional space with a frequency dependent
index of refraction, $n = n(\omega)$, occupying half the spacetime
and vacuum in the other half:
\begin{equation}
n = \left\{ \begin{array}{cl}
                 n(\omega) & \mbox{for } x\leq 0,\\ 
                 1 & \mbox{for }x>0.
              \end{array}\right.
\end{equation}
 
There are two simplifications to the mode structure in two
dimensions. Both result from the lack of spatial dimensions
transverse to the interface.  First there are no evanescent modes, thus
the mode structure is characterized solely by $k_1 = k$ with $\omega
= |k|$.  The positive frequency right-going modes, $k>0$, then take
the form
\begin{equation}
f_k(x) = {1\over{\sqrt{4\pi\omega n}}} \left\{ \Theta(-x) \left[ 
e^{i(nkx-\omega t)} + R e^{-i(nkx+\omega t)} \right] + \Theta(x)
\sqrt{n} T e^{i(kx-\omega t)} \right\}.
\end{equation}
Similarly, the left-going modes, $k<0$, are
\begin{equation}
f_k(x) = {1\over{\sqrt{4\pi\omega}}} \left\{ \Theta(-x)  
{1\over\sqrt{n}} T e ^{i(nkx-\omega t)}  + \Theta(x)\left[
e^{i(kx-\omega t)} - R e^{-i(kx+\omega t)} \right]\right\}.
\end{equation}
The second simplification is that there is just a single set of
reflection and transmission coefficients, 
\begin{equation}
R = { {n-1} \over{n+1} }  \:\:\:\mbox{  and  }\:\:\: 
 T = {{2\sqrt{n}}\over {n+1}},
\end{equation}  
respectively, which satisfy
\begin{equation}
R^2 + T^2 = 1.\label{eq:refl_trans}
\end{equation}

The Wightman Green's function for two points in the vacuum region
of the spacetime is
\begin{eqnarray}
D^+(x,x') & = & \int_{-\infty}^{+\infty} dk \, f^*_{k}(x')\, f_{k}(x) 
,\nonumber\\
&=& {1\over{4\pi}}\left\{ \int_{-\infty}^{0} {dk\over\omega}\, \left[
e^{ik(x-x')} - 2 R \,\cos{k(x+x')} + R^2 \,e^{-ik(x-x')}
\right] e^{-i\omega(t-t')} \right.\nonumber\\
&& + \left.\int_{0}^{+\infty} {dk\over\omega} \,T^2 \,e^{ik(x-x')}
e^{-i\omega(t-t')}\right\}.
\end{eqnarray}
In Minkowski spacetime, the Wightman function is
\begin{equation}
D^+_{\mbox{\tiny Mink.}}(x,x') = {1\over{4\pi}}\int_{-\infty}^{+\infty} 
{dk\over\omega} \, e^{ik(x-x')} e^{-i\omega(t-t')}.
\end{equation}
We can renormalize the Wightman function in the vacuum half space by
subtracting the Minkowski spacetime Wightman function to yield
\begin{eqnarray}
D^+_{\mbox{\tiny Ren.}}(x,x') &=& D^+(x,x') - D^+_{\mbox{\tiny Mink.}}
(x,x') ,\nonumber\\
&=& {1\over{4\pi}}\left\{ \int_{-\infty}^{0} {dk\over\omega}\, \left[
- 2 R \,\cos{k(x+x')} + R^2 \,e^{-ik(x-x')}\right] e^{-i\omega(t-t')}
\right.\nonumber\\
&& - \left.\int_{0}^{+\infty} {dk\over\omega} R^2 \,e^{ik(x-x')}
e^{-i\omega(t-t')}\right\},\nonumber
\end{eqnarray}
where we have made use of Eq.~(\ref{eq:refl_trans}).  This can be
further reduced by making the change of variable in the first
integral of $k = -\omega$, resulting in
\begin{equation}
D^+_{\mbox{\tiny Ren.}}(x,x') = -{1\over{2\pi}} \int_{0}^{\infty}
{d\omega\over\omega}
\,R(\omega)\, \cos{\omega(x+x')}\,e^{-i\omega(t-t')}.\label{eq:D_ren}
\end{equation}

The stress-tensor in the vacuum region for arbitrary coupling is 
\cite{Brl&Dv}
\begin{equation}
T_{\mu\nu} = (1 - 2\xi) \phi_{;\mu} \phi_{;\nu} + (2\xi -{1\over 2})g_{\mu\nu}
g^{\rho\sigma}\phi_{;\rho} \phi_{;\sigma} - 2\xi \phi_{;\mu\nu}\phi,
\end{equation}
where $\xi$ is the coupling constant.
The renormalized vacuum expectation value of the stress-tensor is
found directly from the renormalized Wightman function
\cite{Brl&Dv},
\begin{equation}
\langle 0 | T_{tt} | 0 \rangle _{\mbox{\tiny Ren.}} = \lim_{x'\rightarrow x}
\left[ {1\over 2} \partial_t\partial_{t'} +({1\over 2} - 2\xi)\partial_x
\partial_{x'} - \xi \left(\partial^2_{t}+\partial^2_{t'}\right)\right]\,
 D^+_{\mbox{\tiny Ren.}}(x,x').
\label{eq:Tmunu_definition}
\end{equation}
Special care is taken in (\ref{eq:Tmunu_definition}) to ensure that
the derivative operators are symmetric under interchange of the primed and
unprimed variables.  While this does not affect the calculation of the 
energy density, it has serious effects for the remaining components of the
stress-tensor.  Substituting Eq.~(\ref{eq:D_ren}) yields
\begin{equation}
\langle 0 | T_{tt} | 0 \rangle _{\mbox{\tiny Ren.}} = -{2\xi\over\pi} 
\int_{0}^{\infty}  d\omega \, \omega \,R(\omega)\, \cos{2\omega x}.
\label{eq:integ_expression}
\end{equation}
Similar calculations show the remaining components of the stress-tensor
vanish,
\begin{equation}
\langle 0 | T_{\mu\nu} | 0 \rangle _{\mbox{\tiny Ren.}} =
\langle 0 | T_{tt} | 0 \rangle _{\mbox{\tiny Ren.}}
\left[ \begin{array}{cc} 1 & 0\\ 0 & 0 \end{array}\right].
\label{eq:2D_stressTensor}
\end{equation}
It is interesting that for $\xi=0$, which is both the minimal and
conformal case in two dimension, we have a vanishing stress-tensor.
The form of the stress-tensor given in Eq.~(\ref{eq:2D_stressTensor})
requires this because of the fact that the trace must vanish for
conformal coupling. That is, the only way the stress-tensor can
have this form and have a vanishing trace is for it to be
identically zero for $\xi = 0$.

\subsection{Energy Conditions}

Before we look at specific examples, we examine the general
behavior of the energy conditions.  First we look at the integrated
energy density (total energy) in a constant time hypersurface.  We
define the energy contained within a volume bounded on the left by
the dielectric surface and on the right by an imaginary boundary at
$x=L$ as
\begin{equation}
E(L)  =  \int_0^L \langle 0 | T_{tt} | 0 \rangle dx
 = - {\xi\over\pi}\int_0^\infty R(\omega) \sin(2L\omega) d\omega.
\end{equation} 
This yields the interesting result that in the limit $L\rightarrow\infty$
the total integrated energy is zero. This follows directly from the
Riemann-Lebesgue lemma for the ordinary Fourier integral \cite{Bnd&Or}.
The requirement is that $\int_0^\infty R(\omega) d\omega$ exist, which
in general is true for models where we require
$\lim_{\omega\rightarrow\infty} R(\omega) = 0$. Thus for the 
dielectric mirror, if the integral condition exists then there
must be equal amounts of positive energy and  negative energy.
One can then infer that there is no vacuum energy divergence on
the surface of the dielectric.  It is interesting to note that the
failure of the integral condition on the reflection coefficient
corresponds to divergent vacuum energies, as demonstrated in the
constant index of refraction example below.

Not much can be said about the WEC other than it is necessarily 
violated at some time along a geodesic observer's worldline.
Only by specifying the reflection coefficient can any specific
comments be made.  However, it is remarkable that the
AWEC can be proven in general.  We begin with a geodesic observer
whose world line is given by
\begin{equation}
x^\mu(\tau) = (1-v^2)^{-1/2}\left(\begin{array}{c} \tau\\v\tau\end{array}
\right) \;\mbox{ and }\;    u^\mu(\tau) = (1-v^2)^{-1/2}\left(
\begin{array}{c} 1\\v\end{array}\right),
\end{equation}
where $\tau = [0,\infty)$.  The vacuum energy density integrated along
this worldline is
\begin{equation}
\int_0^\infty \langle T_{\mu\nu} \rangle u^\mu u^\nu d\tau =
-{\xi\over \pi} (1-v^2)^{-1/2}  \lim_{\tau\rightarrow\infty}\int_0^\infty
R(\omega) \sin\left[ 2(1-v^2)^{1/2}\omega v \tau \right],
\end{equation}
which, as we have seen above, is zero for any observer that has $v>0$.
Thus the AWEC is exactly satisfied. This is because such observers
sweep through all of the energy distribution, encountering equal amounts
of positive and negative energy along the way.  For stationary observers,
the AWEC may or may not be satisfied depending on the position at which
the observers sit.  

\subsection{Examples}

\subsubsection{Constant Index, Vacuum and Perfect Reflector}

Let us consider the trivial case where $n(\omega) = \mbox{constant} = n_0 $
for all frequencies.  This makes the reflection coefficient,
\begin{equation}
R(\omega) = R_0 = {{n_0-1}\over{n_0+1}},
\end{equation} independent of frequency as well. Thus the energy density
is
\begin{equation}
\langle 0 | T_{tt} | 0 \rangle _{\mbox{\tiny Ren.}} = -{2\xi\over\pi}\, 
R_0\, \int_{0}^{\infty}  d\omega \, \omega \cos{2\omega x}\, .
\end{equation}
However this integral is not well defined, and requires a regularization
scheme to be evaluated. One method  is to introduce a cutoff function,
carry out the now well defined integral, and then take the limit as the
cutoff goes to one for all frequencies.  An example cutoff function is
$f(\omega) = e^{-\alpha\omega}$ in the limit as $\alpha \rightarrow 0$.
Therefore, the regularization of the vacuum energy results in
\begin{eqnarray}
\langle0 | T_{tt} | 0 \rangle _{\mbox{\tiny Ren.}} &  =& -{2\xi\,R_0
\over\pi}\, \lim_{\alpha\rightarrow 0} \int_{0}^{\infty}  d\omega \,
\omega e^{-\alpha\omega}\cos{2\omega x},  \nonumber\\
&=& {2\xi\,R_0\over\pi}\,\lim_{\alpha\rightarrow 0} {{ (2x)^2 - \alpha^2 } 
\over {\left[(2x)^2 + \alpha^2\right]^2}}, \nonumber\\ 
&=& {\xi\,R_0 \over 2\pi x^2}.\label{eq:simplest}
\end{eqnarray}
The first thing to note is the divergence of the vacuum energy density
at the surface of the dielectric.  The divergence is expected from the
Riemann-Lebesgue lemma,  as the integral of the reflection coefficient
over all frequencies is divergent.  This is identical, up to the factor
of $R_0$, to the known result for the perfectly reflecting mirror.
Actually the index of refraction in this case is only modifying the
numerical coefficient of the functional form associated with the
perfectly reflecting mirror.
Therefore, the replacement of the perfectly reflecting mirror with a
dielectric results in the strength of the vacuum energy density
being governed by the index of refraction.  Low indices yield very
weak vacuum energy densities relative to the perfect reflector, while
high indices yield appreciable fractions of the perfect reflector
vacuum energy.

At one end of the spectrum for Eq.~(\ref{eq:simplest}) is that of
the dielectric having an index of refraction of unity.  This is the
case of the vacuum spacetime everywhere.  In this case the reflection
coefficient vanishes and there is no vacuum energy density. This result
is consistent with that expected from vacuum renormalization in 
two-dimensional Minkowski spacetime.

At the other end of the spectrum is the perfect reflector, which is
defined as the reflection coefficient being unity for all frequencies,
$R(\omega) = 1$. Associated with this is an infinite index of refraction
which is rather unphysical or at least ill defined. Thus, it may be
better the think of the perfect reflector to be the limit of the
reflection coefficient going to one, 
\begin{equation}
\langle 0 | T_{tt} | 0 \rangle _{\mbox{\tiny Perfect Reflector}} = 
{\xi \over 2\pi x^2}.
\end{equation}
This is the standard result for the vacuum energy in a two-dimensional
spacetime with perfectly reflecting boundary conditions

\subsubsection{Discrete Cutoff}

The next example to consider is a dielectric which has a constant index
of refraction up to a cutoff frequency $\omega_c$ and unity thereafter,
\begin{equation}
n(\omega) = \left\{\begin{array}{cc} n_0 & \mbox{for }\omega \leq
\omega_c ,\\  1 & \mbox{for }\omega > \omega_c . \end{array}\right.
\end{equation}
If we insert this into Eq.~(\ref{eq:integ_expression}), we find
\begin{eqnarray}
\langle 0 | T_{tt} | 0 \rangle _{\mbox{\tiny Ren.}} &=& -{2\xi\over\pi}\, 
\left( {{n_0-1} \over{n_0+1}}\right)\, \int_{0}^{\omega_c}  d\omega \,
\omega \cos{2\omega x},\nonumber\\
&=&{2\xi\,{\omega_c}^2\over\pi} \left( {{n_0-1} \over{n_0+1}}\right)
{1\over\chi^2} \left[1 - \cos (\chi) - \chi\sin (\chi)\right],
\label{eq:discrete}
\end{eqnarray}
where $\chi = 2\omega_c\,x$ is a dimensionless quantity. From this
example we see that the reflection coefficient  is acting as the
regulator to obtain finite results from what would otherwise be an
ill defined energy density integral.  Thus more realistic boundary
conditions leads to a stress-tensor that is self-regularized.

\begin{figure}[ht]
\begin{center}
\leavevmode\epsfxsize=5in
\epsfbox{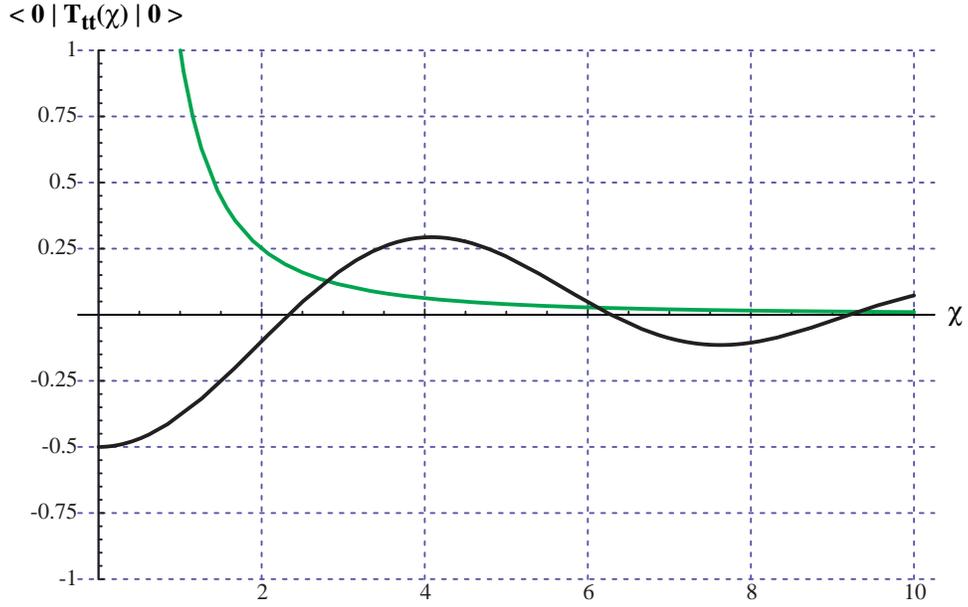}
\end{center}
\caption[Spatial variation of the vacuum energy density for the
discrete cutoff.]
{Spatial variation of the vacuum energy density for the discrete
cutoff in two dimensions. Unlike the perfectly reflecting boundary
vacuum energy, given by the lighter curve, the
energy density for the discrete cutoff does not diverge as one
approaches the interface. Here the energy density is expressed
in units of ${2\xi\,{\omega_c}^2 \over\pi} \left( 
{{n_0-1} \over{n_0+1}}\right)$.}
\label{fig:vacuum1}
\end{figure}

Eq.~(\ref{eq:discrete}) is plotted in Fig.~\ref{fig:vacuum1} where
the most noticeable fact is the disappearance of the divergence in the
vacuum energy density as the dielectric-vacuum interface is approached. 
The divergence has been replaced with a finite value of opposite sign.
From the knowledge of the energy conditions above we know that this is
a general feature of the vacuum energy for frequency dependent indices
of refraction.

\subsubsection{Exponentially Decaying Reflection Coefficient}
\label{sect:2D_decay}

\begin{figure}[hb]
\begin{center}
\leavevmode\epsfxsize=5in
\epsfbox{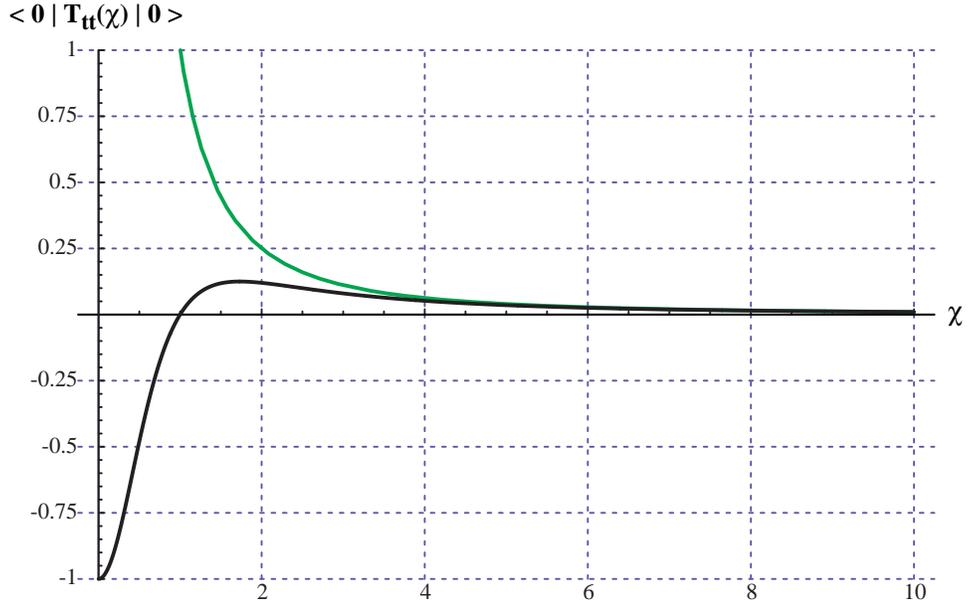}
\end{center}
\caption[]
{Spatial variation of the vacuum energy density for an exponentially
decaying reflection coefficient in two dimensions. Near the dielectric
mirror, the
functional form of the vacuum energy density differs significantly
from that of the perfectly reflecting mirror, given by the lighter
curve.  However, in the large
$\chi$ limit, the dielectric mirror vacuum energy density reproduces
the $\chi^{-2}$ behavior of the perfectly reflecting mirror. The energy
density has been normalized to units of ${2\xi\,{\omega_c}^2 \over\pi}$.}
\label{fig:vacuum2}
\end{figure}

An example of a frequency dependent index of refraction that does not have
a discrete cutoff is
\begin{equation}
n(\omega) = {{1 + e^{-\alpha\omega}}\over{1-e^{-\alpha\omega}}}.
\end{equation}
We define a cutoff frequency $\omega_c = \alpha^{-1}$ as the point where
the reflection coefficient $R(\omega_c) = 1/e$. The associated energy
density is
\begin{eqnarray}
\langle 0 | T_{tt} | 0 \rangle _{\mbox{\tiny Ren.}} &=& -{2\xi\over\pi}\, 
\, \int_{0}^{\infty}  d\omega \, \omega e^{-\alpha\omega}\cos{2\omega x},
\nonumber\\&=&{2\xi\over\pi}\,{\partial\over\partial\alpha} \,\int_{0}^{\infty}
d\omega\, e^{-\alpha\omega} \cos{2\omega x},\nonumber\\
&=& {2\xi\,{\omega_c}^2\over\pi} { {\chi^2 -1} \over
 {\left(\chi^2+1\right)^2}}.
\label{eq:exponential}
\end{eqnarray}
This function is plotted in Fig.~\ref{fig:vacuum2} which clearly
demonstrates the two characteristic behaviors of the vacuum energy
for frequency dependent dielectrics.  The first is that the vacuum
energy asymptotically approaches the form of the perfectly reflecting
vacuum energy at large distances from the mirror.  The second is
the introduction of a new energy density near the mirror surface which
goes to a finite value on the  surface and has the
opposite sign to the perfectly reflecting vacuum energy.  Integrating
Eq.~(\ref{eq:exponential}) over $x=[0,\infty)$ yields zero, confirming
that there is equal positive and negative energy in the vacuum.

\subsubsection{Gaussian Reflection Coefficient}

\begin{figure}[ht]
\begin{center}
\leavevmode\epsfxsize=5in
\epsfbox{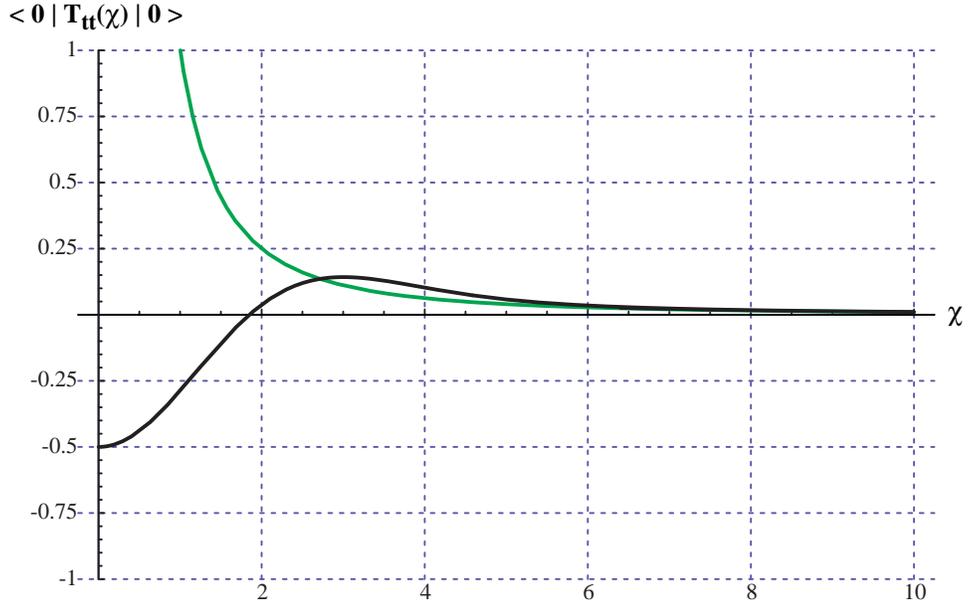}
\end{center}
\caption[]
{Spatial variation of the vacuum energy density for a Gaussian
reflection coefficient in two dimensions.  The energy density is
expressed in units of ${2\xi\,{\omega_c}^2 \over\pi}$.  The vacuum
energy density for a perfect reflector is given by the lighter
curve.}
\label{fig:vacuum3}
\end{figure}

Finally, consider the case where the reflection coefficient
is a Gaussian.  The index of refraction is given by
\begin{equation}
n(\omega) = {{1 + e^{-\alpha^2\omega^2}}\over{1-e^{-\alpha^2\omega^2}}}.
\end{equation}
As was the case in the preceding example, the cutoff frequency is 
defined as $\omega_c = \alpha^{-1}$.  The energy density is easily
calculated,
\begin{eqnarray}
\langle 0 | T_{tt} | 0 \rangle _{\mbox{\tiny Ren.}} &=& -{2\xi\over\pi}\,
\, \int_{0}^{\infty}  d\omega \, \omega e^{-\alpha^2\omega^2}\cos{2\omega x},
\nonumber\\
&=&{2\xi\omega_c^2\over\pi}\,\left\{ -{1\over2} \left[1- \sqrt{\pi}\,
{\chi\over 2} \,e^{-\left({\chi\over2}\right)^2}\,
\mbox{Erfi}\left({\chi\over2}\right)\right]\right\},
\end{eqnarray}
where $\mbox{Erfi}(x) = -i\,\mbox{Erf}(ix)$ is the imaginary error function.
The behavior of the energy density is plotted in Figure~\ref{fig:vacuum3},
and we see that its behavior is quite similar to that for the exponential
reflection coefficient of the preceding example.

\section{Four-dimensional dielectric half-space}

\subsection{Vacuum stress-tensor}

The dielectric to vacuum half space is now considered in four dimensions,
with
\begin{equation}
n = \left\{ \begin{array}{cl}
                 n(\omega) & \mbox{for } x_1=x\leq 0,\\ 
                 1 & \mbox{for }x_1=x>0.
              \end{array}\right.
\end{equation}
 The mode structure has the form of Eqs.~(\ref{eq:Gen_right_mode})
and (\ref{eq:Gen_left_mode}) and their respective transmission and
reflection coefficients as derived in Section \ref{sec:modes}.
The renormalized Wightman function for both spacetime points in
the vacuum region is
\begin{eqnarray}
D^+_{\mbox{\tiny Ren.}}(x,x') &=& {1\over 16\pi^3} \int_{-\infty}^\infty dk_z
\int_{-\infty}^\infty dk_y  \int_{-\infty}^0 {dk_x\over\omega}
\;R_{lg} \left[e^{ik_x(x+x')} +  e^{-ik_x(x+x')} \right] 
e^{i\left[k_y(y-y') + k_z(z-z') -\omega(t-t')\right]} \nonumber\\
&+& {1\over 16\pi^3} \int_{-\infty}^\infty dk_z \int_{-\infty}^\infty dk_y  
\int_{-\infty}^0 {dk_x\over\omega} \; \left(| R_{lg} |^2 - 1\right)
e^{i\left[-k_x(x-x') + k_y(y-y') + k_z(z-z') -\omega(t-t')\right]}
\nonumber\\
&+&{1\over 16\pi^3} \int_{-\infty}^\infty dk_z \int_{-\infty}^\infty dk_y 
\int_0^\infty {dk_x\over\omega} \; n |T_{rg}|^2 e^{i\left[(\beta x -\beta^* x')
+ nk_y(y-y') + nk_z(z-z') -\omega(t-t')\right]}.
\end{eqnarray}
It is a straightforward but tedious task to act on the Wightman function 
with the appropriate derivative operator to find all the components of the
stress-tensor.  For simplicity we will consider the stress-tensor of the
minimally coupled scalar field only.  Three components are found to be
non-zero,
\begin{equation}
\langle 0| T_{\mu\nu} |0 \rangle_{\mbox{\tiny Ren.}} = \mbox{diag} \left[
\langle 0|T_{tt}|0\rangle , 0, \langle0|T_{ll}|0\rangle,
\langle0|T_{ll}|0\rangle\right].
\label{eq:4D_StressTensor}
\end{equation}
The formal expression for each is 
\begin{eqnarray}
\langle 0|T_{tt}|0\rangle &=& {1\over 8\pi^2} \int_0^\infty \omega^3 d\omega
\left[ 2\int_{\pi/2}^\pi d\theta \,\sin^3\theta \,R_{lg} \,\cos\left(2\omega x 
\cos\theta\right) + \right.\nonumber\\
&& \hspace{1in}\left. + n^3 \int_{\sin^{-1}(1/n)}^{\pi/2} d\theta \,\sin^3\theta
\,|T_{rg}|^2 \,e^{-2\omega x \sqrt{n^2 \sin^2\theta - 1}}\right]
\label{eq:E_dens}
\end{eqnarray}
and
\begin{eqnarray}
\langle 0|T_{ll}|0\rangle &=& -{1\over 2}\langle 0|T_{tt}|0\rangle
+{1\over 8\pi^2}\int_0^\infty \omega^3 d\omega \left[ 2\int_{\pi/2}^\pi
d\theta \,\sin\theta 
\,R_{lg} \,\cos\left(2\omega x \cos\theta\right) + \right.\nonumber\\
&& \hspace{2in}\left. + n \int_{\sin^{-1}(1/n)}^{\pi/2} d\theta \,\sin\theta
\,|T_{rg}|^2 \,e^{-2\omega x \sqrt{n^2 \sin^2\theta - 1}}\right]
\label{eq:T_ll_part}
\end{eqnarray}
where we have transformed to spherical momentum coordinates about the
normal to the interface,
\begin{equation}
\left(k_x,\,k_y,\,k_z\right)\rightarrow \left(\omega\cos\theta,\,\omega\sin\theta
\cos\varphi,\,\omega\sin\theta\sin\varphi\right),
\end{equation}
and the integral over $\varphi$ has already been carried out.  The needed
reflection and transmission coefficients now take the form
\begin{equation}
R_{lg} = {{-\cos\theta - \sqrt{n^2-\sin^2\theta}} \over
{-\cos\theta+ \sqrt{ n^2-\sin^2\theta}}}
\hspace{0.5in}\mbox{   and   }\hspace{0.5in}
T_{rg} = {{2\cos\theta} \over
{\cos\theta + i\sqrt{\sin^2\theta -{1\over n^2}}}}.
\label{eq:R_and_T}
\end{equation}

\begin{figure}[hb]
\begin{center}
\leavevmode\epsfxsize=5in
\epsfbox{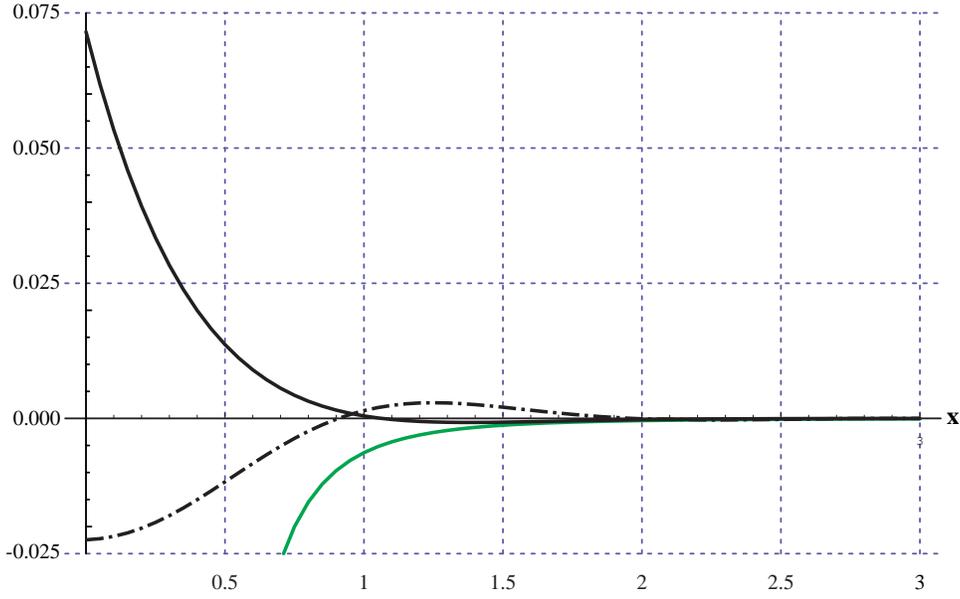}
\end{center}
\caption[]
{Numerical simulation of the $T_{tt}$ and $T_{ll}$ components of the
vacuum stress-tensor, solid and dashed lines respectively,
in the free space region for a planar frequency dependent dielectric to
vacuum interface in four dimensions. The wave equation for the scalar field
has minimal coupling and the model of the index of refraction is 
$n(\omega) = 1 + 3 \exp(-w^2)$. The lighter line is the case of the perfect
reflector.}
\label{fig:4D}
\end{figure}

The integral expressions involving the cosine and left-going
reflection coefficient is the four-dimensional equivalent to that found
previously for the two-dimensional case.  The new term involving the
right-going transmission coefficient is the contribution to the vacuum
polarization due to the evanescent modes.  Recall that evanescent modes
do not exist in two dimensions, so we cannot relate these new terms to
anything seen previously. However in higher dimensions,
their importance to the vacuum stress-tensor cannot be understated.  For
example, in the energy density term, we see from the form of
Eq.~(\ref{eq:E_dens})
that the evanescent modes contribute only positive energy density.  The
components of the stress-tensor are plotted as a function of position
in Figure~\ref{fig:4D} for a specific choice of the index of refraction.

Again we find that the stress-tensor components asymptotically approach
the case of the perfect reflector in the large distance limit.  Close
to the dielectric to vacuum interface, their behavior is again modified
taking on the opposite sign and going to a finite value. In the 
$n\rightarrow\infty$ limit for all $\omega$, the dielectric stress-tensor
reduces to the standard form of the perfect reflecting plate stress-tensor
\cite{Fulling}
\begin{equation}
\langle 0| T_{\mu\nu} |0 \rangle_{\mbox{\tiny Ren.}} = -{1\over 16\pi^2 x^4}
\mbox{diag} \left[1,\,0,\,-1,\,-1\right].
\end{equation}
It should be noted that the perfectly reflecting plate stress-tensor is
a Lorentz tensor that can be formed from the metric $\eta_{\mu\nu}$ and
the unit normal to the plate.  However this is not a property of the
dielectric stress-tensor in Eq.~(\ref{eq:4D_StressTensor}). In general, 
$\langle 0|T_{tt}|0\rangle \neq -\langle0|T_{ll}|0\rangle$ for an
arbitrary choice of the function $n(\omega)$.  The failure of  
$\langle 0| T_{\mu\nu} |0 \rangle_{\mbox{\tiny Ren.}}$ to be a 
Lorentz tensor results from $n(\omega)$ being a frame dependent
quantity. 

\subsection{Energy Conditions}

Again, the mean energy density per unit plate area, the WEC and the AWEC
are considered. We begin with the energy density as seen by a static
observer integrated  along the normal direction to the interface in a
constant time hypersurface.  We evaluate
\begin{eqnarray}
E = \int_0^{+\infty}\ \langle 0|T_{tt}(x)|0\rangle dx &=& {1\over 8\pi^2}
\lim_{x_0 \rightarrow\infty}\int_0^{x_0} dx \left\{ \int_0^{\infty} \omega^3
d\omega \left[ 2\int_{\pi/2}^{\pi}  \,\sin^3\theta\, R_{lg}\, \cos
\left( 2\omega x \cos\theta\right)d\theta\right.\right.+\nonumber\\
&& \left.\left. + n^3 \int_{\sin^{-1}(1/n)}^{\pi/2}
\sin^3\theta \,| T_{rg} |^2 \,e^{-2\omega x \sqrt{n^2 \sin^2\theta - 1}}
d\theta\right]\right\}
\end{eqnarray}
Here $E$ represents the total amount of energy in a rectangular column,
with unit base area at the dielectric to vacuum interface and of infinite
length in the normal direction. Interchanging the orders of integration
in the above expression, then evaluating in $x$ gives
\begin{equation}
E = {1\over 16\pi^2}
\lim_{x_0 \rightarrow\infty} \int_0^{\infty} \omega^2\,
\left( I_1 + I_2 + I_3 \right)d\omega, 
\end{equation}
where $I_1$, $I_2$ and $I_3$ are the three remaining angular integrals
With the aid of Eq.~(\ref{eq:R_and_T}), these integrals can be individually 
considered. The first is
\begin{eqnarray}
I_1 &=& 2 \int_{\pi/2}^\pi {\sin^3\theta\over\cos\theta}
\, R_{lg}\, \sin \left( 2\omega x_0 \cos\theta\right)\,d\theta\nonumber\\
&=& -2\int_0^1 {\sin(2\omega x_0 y)\over y} \,dy +{2\over(n^2-1)}
\left\{2\int_0^1(1-y^2) \sqrt{(n^2-1)+y^2}\sin(2\omega x_0 y)  
\,dy\right.+\nonumber\\
&&\;\;\left.+\int_0^1 \left[2y(y^2-1) + (n^2-1)y\right]\sin(2\omega x_0 y)
\,dy \right\},
\end{eqnarray}
where the change of variable $y = -\cos\theta$ has been made.
At this point we apply the Riemann-Lebesgue lemma to the
second and third terms of the expression when we take the limit
$x_0 \rightarrow \infty$.  It is easy to show that
\begin{equation}
\int_0^1 |(1-y^2) \sqrt{(n^2-1)+y^2}| dy = {1\over 16}\left[
-2n(n^2-3) + (n^4 +2n^2 -3) \ln\left({n+1 \over n-1}\right) \right]
\end{equation}
and
\begin{equation}
2\int_0^1 \left| 2y(y^2-1)+(n^2-1)y\right|dy =
\left\{ \begin{array}{cl}
                 {1\over2}(n^4 -4n^2+5) & \mbox{for } 1\leq n \leq \sqrt{3},\\ 
                 n^2-2 & \mbox{for } n>\sqrt{3}
              \end{array}\right.
\end{equation}
exist for all values of $1\leq n <\infty$.  Therefore, both terms in
$I_1$ vanish, resulting in
\begin{equation}
\lim_{x_0\rightarrow\infty} I_1 = -2 \lim_{x_0\rightarrow\infty}\int_0^1
{\sin(2\omega x_0 y)\over y} \,dy= -2\lim_{x_0\rightarrow\infty} Si(2\omega x_0)
= -\pi,
\end{equation}
where $Si(x)$ is the sine integral.

The next angular integral to be evaluated,
\begin{equation}
I_2 = -n^3 \int_{\sin^{-1}(1/n)}^{\pi/2} {\sin^3\theta\over\sqrt{n^2 
\sin^2\theta - 1}} \,| T_{rg} |^2 \,e^{-2\omega x \sqrt{n^2 \sin^2\theta - 1}}
d\theta,
\end{equation}
has exponential suppression in the large $x_0$ limit.  The only place of
concern would be when $\omega = 0$.  However, the factor of $\omega^2$
in the frequency integral would cancel any contribution from that point.
Therefore the entire integral vanishes in the limit $x_0\rightarrow\infty$. 
An exact analysis using Laplace's method for two-dimensional integrals
to find the asymptotic expansion of this integral for large $x_0$ yields
the same result.

The final integral,
\begin{equation}
I_3 = -n^3 \int_{\sin^{-1}(1/n)}^{\pi/2} {\sin^3\theta\over\sqrt{n^2 
\sin^2\theta - 1}} \,| T_{rg} |^2 \, d\theta = {\pi\over 4} (n^2+3),
\end{equation}
results in a constant that is independent of $x_0$.  When all the angular
integral contributions are added together, the total energy contained
in the column of unit area is
\begin{equation}
E = {1\over 64 \pi}\int_0^{\infty} \omega^2 \,\left[n^2(\omega)-1\right]
\,d\omega.
\label{eq:4D_total}
\end{equation}
Immediately, we see that the energy is always greater than zero for any
model that has $1\leq n <\infty$.  In addition, this term will be finite
for models where $n(\omega)\rightarrow 1$ in the $\omega\rightarrow\infty$
limit.  Therefore, such models will not have surface divergences in the
stress-tensor.

Since the vacuum stress-tensor asymptotically approaches that of the perfect
reflector in the large distance limit, we can assume that the negative energy
region is distant from the mirror, and most likely very small
in overall magnitude.  Because of Eq.~(\ref{eq:4D_total}), the positive energy
must be of significantly greater magnitude and close to the dielectric mirror.
This behavior is evident in Fig.~\ref{fig:4D}.  However, until the frequency
dependence for the index of refraction is given, nothing exact can be said
about the WEC. 

The case of the AWEC is strikingly different.  In fact, the averaged
energy along the worldline of a moving observer with nonzero normal
velocity is always positive definite.  Consider the geodesic
\begin{equation}
x^\mu(\tau) = \gamma \left[ \begin{array}{c} \tau\\ v_x\tau\\ v_y\tau\\
v_z\tau\end{array}\right],\hspace{0.3in}\mbox{where}\;\;\gamma = \left(
1-v_x^2-v_y^2-v_z^2\right)^{-1/2}
\end{equation} 
and  $v_x \neq 0$.  We are assuming that the 
observer is moving outward from the mirror, so $v_x$ is positive.
The energy density averaged along the observer's worldline is
\begin{equation}
\int_0^{\infty} \langle T_{\mu\nu}(\tau)
\rangle_{\mbox{\tiny Ren.}} u^\mu u^\nu d\tau = \gamma^2\lim_{\tau_0\rightarrow\infty}
\int_0^{\tau_0} \left[ \langle 0 | T_{tt} | 0 \rangle + \left(v_y^2+v_z^2
\right) \langle 0 | T_{ll} | 0 \rangle\right] d\tau.
\end{equation}
The first term in the integral is identical, up to factors of $\gamma$
and $v_x$, to that done to arrive at Eq.~(\ref{eq:4D_total}).  The same
analysis using the Riemann-Lebesgue lemma on the second term involving
$\langle 0 | T_{ll} | 0 \rangle$ finds that nonzero contributions come
only from the $-\langle 0 | T_{tt} | 0\rangle$ part when integrating
Eq.~(\ref{eq:T_ll_part}).  Thus, we obtain the AWEC,
\begin{equation}
\int_0^{\infty} \langle T_{\mu\nu}(\tau) \rangle_{\mbox{\tiny Ren.}}
 u^\mu u^\nu d\tau = {1\over 128 \pi} {\gamma\over v_x}\left(2-v_y^2-v_z^2\right)
\int_0^{\infty} \omega^2 \,\left[n^2(\omega)-1\right]\,d\omega \geq 0.
\end{equation}
The AWEC is satisfied for any outward or inward traveling geodesic
observer's worldline.  This results from the observer crossing the
region of positive energy close to the dielectric to vacuum interface.
However, for observers traveling parallel to the interface the AWEC
can fail if the observer is in the negative energy region.  Such 
geodesics never cross the positive energy region of the spacetime.

\section{Summary}

In the preceding sections, we have shown that the general characteristics
of the stress-tensor in the vacuum region outside of a planar dispersive
dielectric can be discussed without specific knowledge of the frequency
dependence of the dielectric.  The divergence of the stress-tensor at
a planar interface is linked to the failure of a set of integrals
involving the reflection and transmission coefficients to exist.  If
such integrals do exist then the mode function expansion of the vacuum
stress-tensor is self regularized.  At sufficiently large distances from
the dielectric mirror, the vacuum stress-tensor asymptotically approaches
that of the perfect reflecting mirror.  However, near the dielectric mirror
the form of the stress-tensor approaches a finite value of opposite sign
to that of the distant vacuum energy.  This leads to the remarkable results
that the mean energy per unit plate area is positive definite (or zero)
and the AWEC is satisfied along all inward or outward half-infinite
timelike geodesics. Taking the geodesic's four-velocity  to unity yields
the ANEC as well.

The results here are for the free quantized scalar field.  However the
mode structure and boundary conditions are identical to that of the transverse
magnetic modes of the electromagnetic field, so we would conjecture there
would be comparable results for the electromagnetic field case.   However
the divergence that would hopefully be removed is not in the stress-tensor
but in the expectation values of the square of the field strengths.   

There is a rich line of future topics to research with respect to the vacuum
stress-tensor with more realistic boundary conditions.  The most immediately
interesting would be to carry out a similar analysis for the quantized
electromagnetic field at a planar dispersive dielectric boundary.  Part of
the foundation for this has already been carried out by Helfer and Lang
\cite{He&La98}.  In addition, a more realistic model for the index of
refraction should be considered that includes conductivity and attenuation.
The study of curved surfaces should also yield interesting results.
One problem that stands out is the Casimir force (and stress-tensor) between
two parallel infinite planar dispersive dielectric half spaces separated
by a gap.  This problem has an extremely interesting mode structure which
includes evanescent, tunneling and waveguide modes.  One can also consider
mixing the dispersive dielectric model with non-zero position uncertainty
for the dielectric \cite{F&SV97b}.  

 \begin{center}{\bf Acknowledgments}\end{center}
The author would like to thank Eric Poisson, L.H. Ford and Donald Marolf
for useful discussions and comments on the manuscript.  This work was
supported by the National Sciences and Engineering Research Council of
Canada.

\newpage
\vfill
\vfill

\end{document}